\documentclass[conference,onecolumn]{IEEEtran}
\usepackage{cite}
\usepackage{amsmath,amssymb,amsfonts}
\usepackage{algorithmic}
\usepackage{graphicx}
\usepackage{textcomp}
\usepackage{xcolor}
\usepackage{listings}
\usepackage{xspace}
\usepackage[colorlinks=false,]{hyperref}
\usepackage{placeins}
\usepackage{tikz}

\usepackage[printwatermark]{xwatermark}
\newwatermark[allpages,color=gray!50,angle=45,scale=2,xpos=0,ypos=0]{Preprint}

\newboolean{showcomments}
\setboolean{showcomments}{true}		
\ifthenelse{\boolean{showcomments}}
{\newcommand{\nb}[2]{
		\fbox{\bfseries\sffamily\scriptsize#1}
		{\sf\small$\blacktriangleright$\textcolor{red}{#2}$\blacktriangleleft$}
	}
}
{\newcommand{\nb}[2]{}
	
}
\usepackage[normalem]{ulem}

\newcommand{\rst}{Restats\xspace}
\newcommand{\rtg}{RestTestGen\xspace}
\newcommand{\rler}{RESTler\xspace}
\newcommand{\bb}{bBOXRT\xspace}
\newcommand{\rets}{RESTest\xspace}
\newcommand{\openapi}{OpenAPI\xspace}
\newcommand{\odg}{ODG\xspace}

\newcommand{\twoxx}{\texttt{2XX}\xspace}
\newcommand{\twooo}{\texttt{200}\xspace}

\newcommand{\fourxx}{\texttt{4XX}\xspace}

\newcommand{\fivexx}{\texttt{5XX}\xspace}

\newcommand{\eg}{e.g.,\ }
\newcommand{\ie}{i.e.,\ }

\renewcommand{\descriptionlabel}[1]{\hspace{\labelsep}\textit{#1}}
\newcommand{\fakeparagraph}[1]{\smallskip\textbf{\textit{#1.~}}}
\usepackage{pifont}
\newcommand{\cmark}{\ding{51}}
\newcommand{\xmark}{\ding{55}}
\newcommand{\tabhighlight}[2][gray!20]{\colorbox{#1}{#2}}

\definecolor{delim}{RGB}{20,105,176}
\definecolor{numb}{RGB}{0, 51, 0}
\definecolor{string}{RGB}{77, 0, 77}

\lstdefinelanguage{json}{
	numbers=left,
	numberstyle=\small,
	frame=single,
	rulecolor=\color{black},
	showspaces=false,
	showtabs=false,
	breaklines=true,
	postbreak=\raisebox{0ex}[0ex][0ex]{\ensuremath{\color{gray}\hookrightarrow\space}},
	breakatwhitespace=true,
	showstringspaces=false,
	basicstyle=\ttfamily\small,
	upquote=true,
	morestring=[b]",
	stringstyle=\color{string},
	literate=
	*{0}{{{\color{numb}0}}}{1}
	{1}{{{\color{numb}1}}}{1}
	{2}{{{\color{numb}2}}}{1}
	{3}{{{\color{numb}3}}}{1}
	{4}{{{\color{numb}4}}}{1}
	{5}{{{\color{numb}5}}}{1}
	{6}{{{\color{numb}6}}}{1}
	{7}{{{\color{numb}7}}}{1}
	{8}{{{\color{numb}8}}}{1}
	{9}{{{\color{numb}9}}}{1}
	{\{}{{{\color{delim}{\{}}}}{1}
	{\}}{{{\color{delim}{\}}}}}{1}
	{[}{{{\color{delim}{[}}}}{1}
	{]}{{{\color{delim}{]}}}}{1},
}

\newcommand\YAMLcolonstyle{\color{red}\ttfamily\small}
\newcommand\YAMLkeystyle{\color{black}\ttfamily\small}
\newcommand\YAMLvaluestyle{\color{blue}\ttfamily\small}

\lstdefinelanguage{yaml}{
	keywords={true,false,null,y,n},
	keywordstyle=\color{darkgray},
	basicstyle=\YAMLkeystyle,                                 
	frame=single,
	numbers=left,
	numberstyle=\small,
	sensitive=false,
	comment=[l]{\#},
	morecomment=[s]{/*}{*/},
	commentstyle=\color{purple},
	stringstyle=\YAMLvaluestyle,
	showstringspaces=false,
	moredelim=[l][\color{orange}]{\&},
	moredelim=[l][\color{magenta}]{*},
	moredelim=**[il][\YAMLcolonstyle{:}\YAMLvaluestyle]{:},   
	morestring=[b]',
	morestring=[b]",
	literate =    {---}{{\ProcessThreeDashes}}3
	{>}{{\textcolor{red}\textgreater}}1
	{|}{{\textcolor{red}\textbar}}1
	{\ -\ }{{\mdseries\ -\ }}3,
}

\begin{document}

\title{Empirical Comparison of Black-box Test Case Generation Tools for RESTful APIs
}

\author{%
	\IEEEauthorblockN{%
		Davide Corradini\IEEEauthorrefmark{1}, 
		Amedeo Zampieri\IEEEauthorrefmark{2}, 
		Michele Pasqua\IEEEauthorrefmark{3} and 
		Mariano Ceccato\IEEEauthorrefmark{4}}
	\IEEEauthorblockA{%
		\textit{Department of Computer Science} \\
		\textit{University of Verona} -- Verona, Italy \\
		Email: %
		\IEEEauthorrefmark{1}davide.corradini@univr.it, 
		\IEEEauthorrefmark{2}amedeo.zampieri@studenti.univr.it, 
		\IEEEauthorrefmark{3}michele.pasqua@univr.it, 
		\IEEEauthorrefmark{4}mariano.ceccato@univr.it}
}

\maketitle

\begin{center}
	\begin{tikzpicture}
	\node (A) at (0,1.25) {Paper accepted for publication in the proceedings of:};
	\node at (0,0.75) {\emph{21$^\text{st}$ IEEE International Working Conference on Source Code Analysis and Manipulation (SCAM 2021)}};
	\node (B) at (0,0) {\footnotesize The present document is the preliminary version of the work prior to peer-review. The final version can be found on the publisher website.};
	\filldraw[rounded corners=2pt,fill=gray,draw=gray!25,opacity=0.25] (B.south west) rectangle (B.east |- 52, 52 |- A.north);
	\end{tikzpicture}
\end{center}

\bigskip

\begin{abstract}
In literature, we can find research tools to automatically generate test cases for RESTful APIs, addressing the specificity of this particular programming domain. However, no direct comparison of these tools is available to guide developers in deciding which tool best fits their REST API project.

In this paper, we present the results of an empirical comparison of automated black-box test case generation approaches for REST APIs. We surveyed the available black-box testing tools that have been proposed in recent literature, finding four usable prototypes: \rtg, \rler, \bb and \rets. We used these tools to generate test cases for 14 real-world REST services. Then, testing results have been analyzed and compared in terms of robustness (\ie success rate) and test coverage. 

Among the considered tools, \rler appears to be the most solid, able to successfully test all case studies (the other tools experienced crashes). Conversely, test cases generated by \rtg scored the highest coverage, suggesting that its testing strategy is the most effective in testing REST~APIs. 
\end{abstract}


\section{Introduction}
\label{sec:intro}


RESTful APIs (or REST APIs for short) are the reference architectural style to design and develop Web APIs, using the REpresentational State Transfer paradigm. They are largely adopted to integrate and interoperate information systems, especially when connected to the cloud~\cite{APIcloud2016}. 

Despite testing being a cornerstone practice in software development to reveal implementation defects, manually writing all test cases for a REST API can be tedious, time-consuming and expensive. Hence, automated test case generation emerged as a way to ease and support developers in testing REST~APIs. 

Automated testing strategies have been proposed and implemented, based on different test case generation algorithms and conceived with different designs to serve different testing purposes. 
Test interactions can be composed, for instance, based on data dependencies among operations,
or applying heuristics to elaborate promising request sequences. 
Moreover, several input data generation techniques have been proposed: either based on documented examples, reusing previously observed values, choosing values from dictionaries, applying mutations, or exploiting constraints among input parameters.


Despite several of them being available, to the best of our knowledge, no guideline is available to help developers in making an informed decision on which tool is more suitable to automatically test a REST API. Indeed, the different testing algorithms implemented by the tools have been designed with different fault model in mind, to expose possibly different faults (\eg data integrity Vs security issues). So, different tools might come with different and incompatible oracles to reveal defects. Thus, a direct comparison on bug detection might~be~unfair.

Considering that automated REST API testing tools typically adopt a black-box approach and base the test case generation process on API interface specifications, an alternative and more fair comparison perspective could be with respect to the interface itself, \ie what extent of the API interface is exercised when running the automatically generated test cases.


This paper defines and adopts a formal framework to empirically compare tools meant to automatically generate test cases for REST APIs. The framework (fully available in our replication package~\cite{ReplicationPackage}) contains 14 open source REST API case studies, available as Docker images to facilitate deployment and replication. 

Four state-of-the-art testing tools have been identified by surveying the relevant literature: \rtg, \rler, \bb and \rets. These tools have been deployed to test the REST APIs case studies under the same conditions, \ie each tool was run with same time budget and the database of each case study was reset before starting each testing session. 
Test cases are compared using eight coverage metrics adopting a black-box viewpoint (proposed by Martin-Lopez et al.~\cite{MartinLopez2019TestCoverage}). 


Experimental results underline that available tools come at different levels of reliability, causing less mature implementations (\eg research prototypes) to crash on some case studies. Indeed, \rler shown to be the most solid tool, since it is the only one applicable to all the case studies. Conversely, \rtg and \bb could work, respectively, on 14 and 8 case studies. \rets, as a research prototype, could be successfully applied only to 2 case studies.

Moreover, results point out that different testing strategies allow maximizing different coverage metrics. In fact, while \bb achieved the highest coverage of parameter values, \rtg achieved the highest coverage on almost all the other metrics, including coverage of paths, parameters, request/response content-types and status codes.



The rest of the paper is organized as follows. Section~\ref{sec:background} covers the background on REST APIs, while Section~\ref{sec:objTools} describes the four state-of-the-art tools for testing REST APIs we employed. In Section~\ref{sec:metrics}, we describe the theoretical measurement framework that we adopted to assess the coverage of the REST APIs testing tools. In Section~\ref{sec:settings}, we describe the experimental setting used to compare the testing tools taken into account. In Section~\ref{sec:results}, we have the results of the comparison. Finally, after discussing related work in Section~\ref{sec:related}, Section~\ref{sec:conclusion} closes the paper.

\section{Background}
\label{sec:background}

\subsection{RESTful APIs}

A RESTful API (or REST API for short) is an API that respects the REST (REpresentational State Transfer) architectural style~\cite{Fielding2000ArchitecturalStyle}. 
REST APIs provide a uniform interface to create, read, update and delete (CRUD) a resource. A resource is generally identified by an HTTP URI, and CRUD operations are usually mapped to the HTTP methods POST, GET, PUT and DELETE.

For example, consider a REST API \emph{PetStore} managing a collection of pets. A possible HTTP URI pointing to the resource could be \texttt{/pets}. In this case, the HTTP operation \texttt{GET pets} is used to retrieve the list of pets and \texttt{POST /pets} could be used to add a new pet to the collection.

The API 
may accept input parameters to specify additional information for executing operations, such as the identifier of the object to retrieve (\eg \texttt{/pets/\{petId\}}) or a structured object to be added to the collection using the POST method.

\subsection{The \openapi Specification}
\openapi\footnote{\url{https://www.openapis.org/}} defines a standard to document REST APIs. According to \openapi, an API service is described using a structured file (either YAML or JSON) that specifies how to reach the API using a URI, which authentication schema is adopted and the details of all the operations available in the API: the input parameters (and their schema) to be used in requests and the schema of responses. 



After an initial header that specifies versions, licenses and the base URL of the API, an \openapi specification contains an array of \emph{paths}, namely the list of URL paths available in the API. In the \textit{PetStore} example we have two paths: \texttt{/pets} and \texttt{/pets/\{petId\}}. Each path supports one or more HTTP methods. Operations in the API are pairs of paths and methods, and usually are identified by an \emph{operation ID}. For instance, the method GET in \texttt{/pets} (\texttt{getPets}) is used to retrieve the list of all the pets, while the method GET in \texttt{/pets/\{petId\}} refers to the operation \texttt{getPetById}, meant to retrieve the \texttt{Pet} object that matches a specific \texttt{petId}. Path parameters are specified directly in the path URL using curly braces, such as the \texttt{petId} input parameter in the previous example.


Request input and output are associated with a \emph{schema} that specifies their type and, optionally, a set of constraints on values (\eg a \emph{min} or \emph{max} value for numeric parameters). Types can be atomic (\eg integers and strings) or structured (\ie compound objects). For instance, the parameter \texttt{petId} of \texttt{/pets/\{petId\}} could be of type \texttt{string}, while the response to the corresponding GET operation is expected to be in JSON, according to the \texttt{Pet} schema (also defined in the specification). A specification not only describes the response format in nominal cases (\eg a response status code \twooo), but it also describes the format expected when errors occur.

\section{Object Tools}
\label{sec:objTools}


In the last years, the research community proposed several approaches to automatic generation of test cases for REST APIs. We surveyed the literature, selecting the available state-of-the-art tools. In particular, we carefully checked the works published in the top testing and software engineering conferences\footnote{ICSE, ESEC/FSE, CCS, ISSTA, ICST, AST, EDOC, ICICT and ICCSDET.} (together with their satellite workshops) and journals\footnote{TOSEM, TSE, EMSE and TOSC.} appeared in the last 4 years. Furthermore, we also used publications search engines (like \emph{IEEE Xplore}\footnote{\url{https://ieeexplore.ieee.org}}) with the following keywords: ``REST'', ``RESTful API'' and ``black-box testing''. In our search, we looked for approaches complying with the following requirements: (i) approaches operating with a black-box perspective; (ii) approaches implemented into a software tool; (iii) approaches with an open-source implementation or with publicly available binaries.

Eventually, we obtained our final set consisting of four tools: \rtg, \rler, \bb, and \rets. All the selected tools have a common characteristic: they use the information documented inside an \openapi specification to build a testing strategy according to their own algorithm. Approaches differ in terms of operation sequence(s) assembly and input values generation. Table~\ref{tab:tools-summary} summarizes the strategies of the four object tools. 

In the following, a brief overview of the peculiarities of the selected tools.


\begin{table}[t]
	\centering
	\caption{Object tools strategies.}
	\label{tab:tools-summary}
	\resizebox{.6\linewidth}{!}{
		{\renewcommand{\arraystretch}{1.2}
		\begin{tabular}{l|l|l}
			\hline
			\textbf{Tool} & \textbf{Operations sequence(s)} & \textbf{Input values} \\
			\hline\hline
			\rtg~\cite{Viglianisi2020RESTTESTGEN} & Data dependencies & \begin{tabular}[c]{@{}l@{}} Random \\ Documented examples \\ Observed values \\ Dictionary \\ Mutations (3) \end{tabular} \\
			\hline
			\rler~\cite{Atlidakis2019RESTler} & \begin{tabular}[c]{@{}l@{}} Data dependencies \\ Full enumeration of sequences \end{tabular} & \begin{tabular}[c]{@{}l@{}} Observed values \\ Dictionary \end{tabular} \\
			\hline 
			\bb~\cite{Laranjeiro2021BBOXRT} & (not reported) & \begin{tabular}[c]{@{}l@{}} Random \\ Mutations (57) \end{tabular} \\
			\hline
			\rets~\cite{MartinLopez2020RESTest} & \begin{tabular}[c]{@{}l@{}} (not reported) \end{tabular} & \begin{tabular}[c]{@{}l@{}} Random$^\ast$ \\ Documented examples$^\ast$ \\ Dictionary$^\ast$ \\ Mutations$^\ast$ (4) \end{tabular} \\
			\hline
		\end{tabular}
		}
	}

	\vspace*{3pt}
	
	\scriptsize{$\ast$ exploiting inter-parameter dependencies}
	
\end{table}

\subsection{\rtg} 
\rtg is an automated black-box test case generation tool for REST APIs proposed by Viglianisi et. al~\cite{Viglianisi2020RESTTESTGEN}. It is Java-based and the executable JARs are available on GitHub\footnote{\url{https://github.com/resttestgenicst2020/submission_icst2020}}.

The strategy of \rtg is based on the \textit{Operation Dependency Graph} (\odg), a graph which encodes data dependencies among the operations available in the API. For instance, the operation \texttt{GET /pets/\{petId\}} (from the example in Section~\ref{sec:background}) depends on the operation \texttt{GET /pets}, that returns a list of valid pet IDs, because the output of the latter can be used as the input for the former. Dependencies in the \odg are inferred from the \openapi specification by matching parameters names and schemas. The \odg is used to optimize the operations testing order, prioritizing operations with satisfied data dependencies. It is updated at runtime according to the results of executed tests. 

\rtg is composed by two modules, the \textit{Nominal Tester} and the \textit{Error Tester}, that are in charge of generating nominal and error test cases, respectively.

Nominal test cases are meant to test nominal interactions with the API: they are generated to comply with the interface documented in the \openapi specification. Previously observed values, if available, are re-used as input values. Alternatively, \rtg uses a dictionary, examples or random values. An oracle, based on the status code of responses, evaluates the outcome of each test case execution: \texttt{2XX} and \texttt{4XX} status codes are classified as successful executions (indeed, those status codes represent correct executions or graceful errors in the HTTP protocol); \texttt{5XX} status codes are instead classified as failures (server errors in the HTTP protocol).

Error test cases are generated by mutating successful nominal test cases. \rtg applies three different mutations: \textit{missing required}, by removing parameters that are documented as mandatory; \textit{wrong input type}, by modifying types of input parameters; and \textit{constraint violation}, by setting unsupported parameter values according to the documented constraints. An oracle, based on the status code of responses, evaluates test cases executions: \texttt{4XX} status codes are classified as successful, meaning that the server correctly identified the malformed input; \texttt{2XX} status codes are classified as failures, because the server accepted a malformed input as valid; and \texttt{5XX} status codes are once again classified as failures, meaning that the malformed input caused a server-side error.

Both nominal and error test cases are also evaluated by a further oracle, which validates the response schema documented in the \openapi specification against the one obtained in the test cases responses. A test case passes if the response matches its schema definition.


\subsection{\rler} 
\rler is a stateful REST API fuzzer presented by Atlidakis et al.~\cite{Atlidakis2019RESTler} at Microsoft Research, written in Python and available on GitHub\footnote{\url{https://github.com/microsoft/restler-fuzzer} (cloned on Dec. 27th, 2020).}.

\rler generates stateful sequences of requests by inferring producer-consumer relations between request types described in the specification. It also dynamically analyzes responses to intelligently build request sequences and avoiding sequences leading to errors.



\rler relies on different test generation algorithms, and each one implements a different test space search logic. The \emph{BFS} algorithm appends every possible compatible request to every existing sequence, namely performing an exhaustive search. In the \emph{BFS-Fast} algorithm, every request is appended to at most one existing sequence. This results in a smaller set than the full BFS approach, but it does not guarantee that every possible request sequence is generated. \emph{BFS-Cheap}~\cite{Atlidakis2020SecurityProperties} implements the dual trade-off of BFS-Fast: all the sequences are generated but only at most two sets of parameters values
(one valid, one invalid) are allowed for each request. Finally, the \emph{Random-walk} algorithm randomly selects a valid request sequence to which append a random request.


To fuzz input values, \rler relies on a user-configurable dictionary. The user can manually extend this dictionary with custom values that better fit the service, or that are known to be more effective in the testing phase. When \rler detects data dependencies among operations, also previously observed values are used as parameters. For error detection, \rler uses HTTP status codes: if a status belonging to the \fivexx class is detected, then the test sequence could have discovered a bug, so it is logged for further analysis.


\subsection{\bb}
\bb is a black-box robustness testing tool for RESTful APIs proposed by Laranjeiro et al.~\cite{Laranjeiro2021BBOXRT}, written in Java and available on the authors' website\footnote{\url{https://eden.dei.uc.pt/~cnl/papers/2020-access.zip}}. The aim of \bb is to assess the robustness of REST APIs observing the behavior of services under test when providing invalid requests.

The peculiarity of \bb is the large number of supported mutations. The provided fault model consists of 57 different mutations applicable to input parameters of various types (numbers, strings, booleans, dates, times, arrays, etc.).

The \bb execution starts with the analysis of the \openapi specification to collect information about the service under test. Subsequently, the \textit{Workload generator} component starts generating and executing valid requests with the aim to understand the behavior of the service under test in absence of faulty workloads. Parameter values are randomly generated to comply with the specification. Requests triggering a \texttt{2XX} response are stored for future use, while interactions that triggered \texttt{4XX} and \texttt{5XX} status codes are either retried with different values or discarded. Authors do not explain the strategy they adopted to order operations.

The next component, the \textit{Faultload generator}, creates faulty requests by mutating successful requests. It applies mutation rules to parameters, one at a time. Faulty interactions are stored for further analysis by test engineers. According to the authors, \bb does not fully automate the analysis of the test cases outcome, so manual intervention is still required.


\subsection{\rets}

\rets is an automated black-box testing tool for RESTful APIs proposed by Martin-Lopez et al.~\cite{MartinLopez2020RESTest}. It is written in Java and the source code is available on a GitHub repository\footnote{\url{https://github.com/isa-group/RESTest/} (cloned on Dec. 27th, 2020).}.

The peculiarity of this tool is the \textit{inter-parameter dependencies} support. Some REST APIs, in fact, impose constraints that restrict not only input values, but also the way in which input values can be combined to fill valid requests. For example, the YouTube API search operation requires the \texttt{publishedAfter} parameter to be greater or equal to \texttt{publishedBefore}. Currently, the \openapi grammar does not support a formal documentation of such dependencies, so Martin-Lopez et al.~\cite{MartinLopez2019CatalogueInterParamDependencies,MartinLopez2021DependenciesAPI} proposed a domain-specific language, called IDL (inter-parameter dependency language), to this aim. \rets uses the documented dependencies to code a constraint satisfaction problem and deploys a reasoner to generate test cases accordingly. Since, at the moment, OpenAPI specifications do not support inter-parameter dependencies, no REST API comes with a dependencies-enriched specification. 
For this reason, the IDL module of \rets was not active in our experiment.

\rets can generate both nominal and faulty test cases using two strategies: (i) random testing (RT), by generating random input values; and (ii) constraint-based testing (CBT), by exploiting constraints of inter-parameter dependencies. Test cases are generated in both settings from a \textit{test model} derived from the \openapi specification. Nominal test cases aim to stress the service with valid inputs to check its behavior against the specification. Faulty test cases derive from nominal test cases by applying mutations (excluding mandatory parameters, using out-of-range values, and violating the JSON schema). Additionally, \rets can generate faulty test cases by violating inter-parameter dependencies. Authors do not explain the strategy adopted to sort operations during testing.

To classify test outcomes, 5 oracles are deployed: (i) status code must be lower than \texttt{500}; (ii) a response must conform the documented schema; (iii) if the request violates one or more parameter specifications, the status code must be different from \texttt{2XX}; (iv) if a request violates inter-parameter dependencies, the status code must be different from \texttt{2XX}; (v) if the request is valid according to the specification and inter-parameter dependencies are met, the status code must be different from~\texttt{4XX}.


\subsection{Other discarded tools}
Initially, we considered other tools than \rtg, \rler, \bb, and \rets, but, for various reasons, they have been excluded from the object tools list. QuickREST, proposed by Karlsson et al.~\cite{Karlsson2020QuickREST}, is a proof-of-concept tool that has not been released as a generic tool. It is, instead, available only as a customized build to work on the case studies of a replication package, in a package-specific version. Hence, it is not possible to apply QuickREST to test APIs other than those from their replication package. Another discarded tool was proposed by Ed-Douibi et al.~\cite{EdDouibi2018AutomaticTestsGen}. In this case, the proof-of-concept tool was developed as plug-in for \emph{Eclipse}. This tool was discarded due to the presence of errors in the source code that prevented it from installing properly. Finally, EvoMaster~\cite{Arcuri2019EvoMaster} has been excluded because, at the time of writing, it does not follow a black-box approach: it requires the availability of the Java source code to perform static and dynamic analysis.

\section{Test Coverage Metrics}
\label{sec:metrics}

When it comes to comparing REST API testing tools, there is not a standard fault model adopted by the state-of-the-art approaches, which might come with different and incompatible oracles to reveal
defects. In this scenario, a direct comparison on bug detection might be
unfair. Hence, to objectively compare automated test case generation tools, we need a methodology to measure the \emph{coverage} of their~test~cases.
 
The four black-box tools that we are comparing assume that the source code of REST APIs is not available. So, source code coverage cannot be the metric for the comparison. An alternative approach is represented by \emph{interface coverage}, which measures the testing coverage with respect to the specification of the REST API rather than to its actual code. This is also motivated by the fact that the REST API testing tools base the test case generation process on the \openapi specification of the service under test.

In this respect, Martin-Lopez et al.~\cite{MartinLopez2019TestCoverage} proposed a test coverage framework based on the API interface description available within the \openapi specification. They introduced ten \textit{coverage metrics} 
to measure the coverage of a test suite as the ratio of the tested elements on to the total number of elements available in the API. 

Although the available measurement framework provides a starting point for an empirical comparison, adaptations are required to turn metrics operative. Indeed, during the empirical adoption of the framework, we realized that some metric definitions were too abstract to be properly applied in practice.

In the following, we present an overview of the metrics proposed by Martin-Lopez et al.~\cite{MartinLopez2019TestCoverage}, along with our adaptations. They are six metrics related to the generated inputs, and four metrics related to the triggered outputs.

\subsection{Input coverage metrics}
Six metrics, called \textit{input coverage metrics}, are meant to measure the capabilities of a test suite \textit{requests} to exercise different parts of the REST API under test. 

\subsubsection{Path coverage} it measures the capability of a test suite to exercise the API paths. It is the ratio of the number of tested paths to the total number of paths documented in the \openapi specification. A test suite reaches 100\% path coverage if its tests send at least one request directed to each path of the API.

\subsubsection{Operation coverage} it measures the capability of a test suite to execute the available operations. It is the ratio of the number of tested operations to the total number of operations described in the \openapi specification. A test suite reaches 100\% operation coverage if there exists at least one request directed to each path with all the documented HTTP methods.

\subsubsection{Parameter coverage} it measures the capability of a test suite to sample all the available parameters on operations. It is the ratio of the number of input parameters used by test cases to the total number of parameters documented in the \openapi specification. A test suite reaches 100\% parameter coverage if all input parameters of all operations are included in requests at least once.

\subsubsection{Parameter value coverage} it measures the capability of a test suite to choose meaningful values for input parameters. It is the ratio of the number of the exercised parameter values to the total number of possible values that parameters can assume according to the \openapi specification. This metric only applies to domain-limited parameters, such as boolean and enum types. A test suite reaches 100\% parameter value coverage if requests contain all the possible values for each parameter of each operation.

\subsubsection{Request content-type coverage} it measures the capability of a test suite to feed endpoints with request bodies of different content-type formats. It is the ratio of the number of tested content-types to the total number of accepted content-types as documented in the \openapi specification. A test suite reaches 100\% request content-type coverage if there exists at least a test request for each accepted content-type. The original definition by Martin-Lopez et al.~\cite{MartinLopez2019TestCoverage} does not consider scenarios of content-types with wildcards (\eg \texttt{application/*}). Such cases turn the number of accepted content-types unbounded. In our adaptation, we assume that request content-type coverage can be computed only when operations content-types have no wildcards, in order for the metric value to be meaningful.

\subsubsection{Operation flow coverage} it measures the capability of a test suite to apply different sequences of operations. It is defined as the ratio of the number of tested flows to the total number of meaningful flows, according to the application business logic. However, as also acknowledged by Martin-Lopez at al.~\cite{MartinLopez2019TestCoverage}, there is no standard definition of what are the meaningful flows for a REST API, and there is no way to document flows in the \openapi specification. Thus, considering that the definition of this metric is not operative, we decided to not include it in our framework.

\subsection{Output coverage metrics}
Four metrics are meant to measure the coverage of a test suite according to \textit{responses} received from the REST API under test. These metrics are called \textit{output coverage metrics}.

\subsubsection{Status code class coverage} it measures the capability of a test suite to trigger responses with \textit{correct} and \textit{erroneous} status code classes. The \openapi specification does not provide primitives to formally document correct or erroneous status code classes. According to the metric definition by Martin-Lopez et al.~\cite{MartinLopez2019TestCoverage}, it is up to the test engineer to define which status codes belong to the \textit{correct} class, and those belonging to the \textit{erroneous} class, based on the semantic of the target API. To maintain a black-box point of view that assumes no knowledge about the semantic of the API under test, we consider the standard semantic provided by the HTTP protocol, \ie the \twoxx class represents a correct execution and \fourxx and \fivexx classes represent an erroneous execution. 
A test suite reaches 100\% status code class coverage when it is able to trigger both correct and erroneous status codes. Conversely, if it only triggers status codes belonging to the same class (either correct or erroneous), the reached coverage is 50\%.

\subsubsection{Status code coverage} it measures the capability of a test suite to trigger responses with different status codes. It is the ratio of the number of obtained status codes to the total number of status codes documented in the \openapi specification, for each operation. A test suite reaches 100\% status code coverage if, for each operation, it is able to test all the status codes.

\subsubsection{Response body properties coverage} it measures the capability of a test suite to trigger responses containing all the properties defined in their schema. A property is, for instance, a key-value pair of a JSON object. This metric is computed as the ratio of the number of obtained properties to the total number of properties defined in the \openapi specification schemas. A test suite reaches 100\% response body properties coverage if it is able to trigger responses whose bodies contain all properties for all response objects. Parsing the response header is not enough to compute this metric and, in addition, the response bodies have to be parsed with different grammars according to the body content-type (e.g., JSON or XML). For this reason, we decided to skip this metric in this work, 
and focus on the other metrics, whose computation is less complex. However, we plan to implement also this metric as future work.

\subsubsection{Response content-type coverage} it measures the capability of a test suite to trigger responses whose body covers different formats. It is the ratio of the number of obtained content-types to the total number of response content-types documented in the \openapi specification. A test suite reaches 100\% response content-type coverage if there exists at least one test response whose body matches each documented content-type, for each operation. Similarly to the request content-type coverage metric, we will compute this metric only when specific content-types are defined with no wildcard.




\subsection{Automatic metrics computation}\label{ssec:reststat}
To automatically compute the coverage metrics 
achieved by the object tools, we have developed \rst~\cite{reststat2021}, 
a Python tool that implements the aforementioned measurement framework. 
Our implementation is tool-agnostic: it does not employ tool-specific log files to compute the coverage. Quite the opposite, it operates by reading a generic HTTP traffic log, composed by request-response pairs. In our validation, to log requests and responses we have routed all the HTTP traffic through a proxy before executing the object tools.

\rst reads the HTTP log and the \openapi specification of the target API, then it computes path, operation, parameter, parameter value, and status code coverage as originally defined by Martin-Lopez et al.~\cite{MartinLopez2019TestCoverage}. It computes input and output content-type coverage, and status code class coverage according to our adaptation, as explained in the previous paragraphs.

\section{Experimental Settings}
\label{sec:settings}

In this section, we provide an experimental evaluation of the performance of the state-of-the-art testing tools for REST APIs. In particular, our aim is to assess the capability of a tool to test real-world case studies, and to measure how effective the generated tests suites are. The complete package to replicate our experiment is available online~\cite{ReplicationPackage}.

\subsection{Research Questions}



Automated black-box RESTful APIs testing approaches are available as research prototypes and, thus, might not be as robust as commercial tools. They might fail or crash with certain API implementations. Before deciding which tool to adopt, a developer might be interested in knowing their maturity and solidity, so the first research question is meant to compare tools with respect to their ability to manage many real-world case studies.

\begin{description}
	\item[\textbf{RQ1}:] How \emph{robust} are automated RESTful APIs test-case generation tools?
\end{description}

The extent of a REST API that can be tested by automated tools is the other important consideration when deciding which testing tool to adopt. So, the second research question is intended to compare tools with respect to the coverage that their test cases can achieve.


\begin{description}
	\item[\textbf{RQ2}:] What is the \emph{coverage} of the test suites emitted by automated RESTful APIs test-case generation tools?
\end{description}

Our empirical investigation will be designed to answer these research questions.

\subsection{Metrics}


Our empirical evaluation of the performance of REST APIs testing tools is based on two different dimensions. First, we consider \emph{robustness}, aiming at assessing to which extent a tool is ready to be effectively usable. This translates to checking how many APIs a tool is able to test out-of-the-box without unexpected errors. 

Second, we consider \emph{coverage}, aiming at assessing the adequacy of the test cases generated by the tools. Considering that all the tools start from the definition of the API interface (input, output and operations) and require no source code access, coverage will be computed with the same viewpoint.
In particular, interface coverage means how much of the behavior of an API, as described in the specification, is tested by the tools. Coverage will be computed using the coverage metrics introduced in Section~\ref{sec:metrics}. 


\subsection{REST APIs Case Studies}
\label{ssec:case-studies}

\begin{table}[t]
	\centering
	\caption{List of the selected case studies.}
	\label{tab:csSummary}
	\resizebox{.6\linewidth}{!}{
		\begin{tabular}{l|c|c|c|c|r}
			\hline
			\textbf{Case Study} & \textbf{Language} & \textbf{Framework} & \textbf{Endpoints} & \textbf{Operations} & \textbf{\# of lines} \\
			\hline\hline
			\texttt{01-Slim} & PHP & Slim & 9 & 18 & 8,566 \\
			\texttt{02-Airline} & Java & Spring Boot & 12 & 30 & 3,859 \\
			\texttt{03-Streaming} & Java & Spring Boot & 5 & 5 & 1,780 \\
			\texttt{04-Petclinic} & Java & Spring Boot & 17 & 47 & 8,550 \\
			\texttt{05-Toggle} & ASP.NET & .NET Core & 8 & 16 & 2,363 \\
			\texttt{06-Problems} & Java & Spring Boot & 5 & 9 & 2,174 \\
			\texttt{07-Products} & Java & Spring Boot & 6 & 14 & 3,451 \\
			\texttt{08-Widgets} & Go & - & 4 & 14 & 1,370 \\
			\texttt{09-Safrs} & Python & Flask & 6 & 18 & 2,787 \\
			\texttt{10-Realworld} & PHP & Laravel & 11 & 19 & 5,278 \\
			\texttt{11-Crud} & Node.js & Express & 1 & 4 & 5,106 \\
			\texttt{12-Order} & PHP & Laravel & 2 & 3 & 3,359 \\
			\texttt{13-Users} & TypeScript & Express & 2 & 5 & 805 \\
			\texttt{14-Scheduler} & Node.js & Express & 26 & 40 & 24,044 \\
			\hline
		\end{tabular}
	}
	\vspace*{-10pt}
\end{table}

For the comparison to be fair, object tools should operate on the same REST APIs, with the same initial conditions. Many publicly hosted APIs are available for free (such as those on \emph{APIs.guru}\footnote{\url{https://apis.guru/browse-apis/}}) and they have been used as case studies for assessing automated testing tools~\cite{Viglianisi2020RESTTESTGEN}. However, they are not appropriate for a direct comparison among several tools, because the state of these APIs can be changed by previous executions of testing tools or by other users accessing them. Hence, different testing tools might work with APIs at different starting state and this might affect the tool performance and, consequently, threaten the validity of our results.

To overcome this state interference problem, we opted for case studies that we can download and run in a controlled local environment. 
To this aim, we searched for REST API implementations among the open-source projects on \emph{GitHub}. 
With local instances of REST APIs, we can set a common starting point for the underlying database and restore a common initial state of the API before starting each testing iteration. 

We started our search with the query strings ``REST'', ``RESTful API'', ``\openapi'' and ``Swagger'', to have an initial list of candidate case studies. Subsequently, we also added query strings that represent framework commonly used to implement REST APIs, such as ``swagger-ui'', ``SpringFox'', ``swagger-jsdoc'', ``flask-swagger''.
Among these APIs, we kept those containing an \openapi specification, because black-box testing tools require it as input. Some services contain this specification directly in the project sources; for some others, the specification is not in the source code, but it can be automatically generated when their underlying frameworks support this feature. This is the case for some services implemented, for instance, using \emph{Spring}~\cite{Spring} or \emph{Flask}~\cite{Flask}.


These potential case studies have been downloaded, compiled and run to discard those that failed either in compiling or in running. 
After this last filtering, our final set of case studies consists of 14 REST APIs, for a total of 114 endpoints and 234 operations (more information in Table~\ref{tab:csSummary}). We consider these APIs as representative of real-world REST APIs because: (i) they are written in different programming languages (PHP, Java, Go, ASP.NET, Python, JavaScript and TypeScript); (ii) they are based on different frameworks and DBMSs; and (iii) they have different levels of complexity in terms of number of operations and dependencies. Applications are mostly query-intensive: their goal is to manage, for instance, an airline, restaurant orders, users, a library, a pet clinic, etc. Some APIs have many dependencies among operations (\eg the airline management system with airports, planes, flights and routes), while others are simpler.

Among the 14 selected working case studies, 7 of them contained small errors in the specification, resulting invalid according to the official \emph{Swagger Editor}\cite{Editor}. Considering that all testing tools expect a valid specification, we manually fixed these errors, paying attention to not alter the intended semantic. Indeed, we applied only minor changes for evident mistakes trivial to solve, such as removing non RFC3986-compliant characters from URLs, changing wrong syntax when \openapi version~2 (Swagger) syntax was used in \openapi version~3 files, moving fields in the right position when they were misplaced, and renaming operations with reused names when they were supposed to adopt unique naming.

\subsection{Experimental Procedure}

In order for the case studies to be testable, it might be necessary to initialize their state with some data. For instance, a ``delete item'' feature can be tested only when the ``item'' data does exist.  Most of the case studies already came with a pre-initialized database, or with a procedure that fills it after installation. The initial database was completely empty only on few cases, so we adopted the following procedure to fill it. We manually interacted with these APIs, executing each documented feature at least once, thus providing some data. After the API has been moved to a testable state, we took a snapshot of the database, representing the initial state to be set when cleaning side effects produced during testing. 

Note that, very few REST APIs provide a sandbox for testing purposes and, even if provided, a sandbox usually does not come with a full-reset mechanism. For this reason, we have created a custom sandbox for each REST API case study, encapsulated into an independent \emph{Docker} container. In this way, we could isolate every service environment, making each test independent of the others. Furthermore, the use of containers allows us to easily restore the same starting point before running each testing tool. 

Each testing tool has been configured with its default settings or, when available, with the settings that their authors deem the most effective in the corresponding paper. In particular, for \rtg and \bb we used the default settings; for \rler we set \textit{BFS-Cheap} as test generation algorithm because showed as the best performer with reduced time budgets; for \rets we used the CBT generator, although without providing any inter-parameter dependency constraint.
Testing tools have been run on each case study with a time budget of 
10 minutes. 
After each run, the case study database has been reset, to start each testing session from a clean baseline state. Test case generation has been repeated 10 times for each case study to control random variation of non-deterministic algorithms integrated in testing strategies. 
The execution log has been captured by a proxy and coverage metrics have been computed by \rst (see Section~\ref{ssec:reststat}).

The experiment has been conducted on an Ubuntu 20.04 desktop computer equipped with an AMD\textsuperscript{\textregistered} FX\textsuperscript{TM}-6300 six core CPU running at 3.5GHz and 16GB of primary memory.

\subsection{Threats to Validity}


We identified the following limitations as potential threats to the validity of our empirical results.

\fakeparagraph{Conclusion validity} In order to draw correct conclusions, the measurements must be reliable.
To limit this threat, we adopted an existing measurement framework, originally proposed by Martin-Lopez et al.~\cite{MartinLopez2019TestCoverage}, with only minimal changes to turn it operative.

\fakeparagraph{Internal validity} To limit external factors that might influence our observations, case studies have been run locally, so that only testing tools could access them. No other end-users could access the case studies during the experiment and alter their state. Moreover, to give all the testing tools the same starting conditions, case studies databases have been reset before each testing session, thus canceling the footprint of previous executions. To make sure that measurements did not influence the testing results, testing tools have not been instrumented. Instead, coverage metrics have been computed by an external measurement tool, 
that just monitors the network traffic between case studies and testing~tools.

Despite each testing tool reporting some kind of coverage statistics, there could be differences among the way these statistics are collected on different tools. To compare consistent data, coverage reported by testing tools have been ignored and 
coverage has been computed by a measurement tool contributed by us.

\fakeparagraph{Construct validity} Considering that testing tools contain non-deterministic components, by chance rare events may have influenced our results. To limit this threat, we measured 10 independent runs and the average coverage has been reported.


\fakeparagraph{External validity} Although we have sampled real REST APIs in our experimental validation, written in different programming languages and with different frameworks, we cannot assume that our results hold for any other arbitrary REST API. Additional experiments on new case studies are needed to corroborate our findings.

\section{Experimental Results}
\label{sec:results}

This section presents the results of our experimental validation to compare REST APIs testing tools according to robustness and coverage, respectively.

\subsection{RQ1: Analysis of Robustness}

All the tools have been run on the same 14 case studies, monitoring crashes and failures. We tried our best to make tools work on all case studies, sometimes even contacting the authors in order to understand the possible reasons for the failures.
Table~\ref{tab:cstested} reports successful executions with tick-marks (\cmark) and failing executions with cross-marks (\xmark). Eventually, in the last line, the table reports the total number of successfully tested case studies for each automated testing tool. 

\begin{table}[t]
	\parbox[t]{.48\linewidth}{
	\caption{Robustness: case studies successfully tested by each tool.} 
	\label{tab:cstested}
	\begin{tabular}{l|cccc}
		\hline
		\textbf{Case study} & \textbf{\rtg} & \textbf{\rler}& \textbf{\bb} & \textbf{\rets} \\ 
		\hline\hline
		\texttt{01-Slim} & \cmark & \cmark & \xmark & \cmark \\
		\texttt{02-Airline} & \xmark & \cmark & \xmark & \xmark \\
		\texttt{03-Streaming} & \xmark & \cmark & \xmark & \xmark \\
		\texttt{04-Petclinic} & \cmark & \cmark & \xmark & \xmark \\
		\texttt{05-Toggle} & \cmark & \cmark & \cmark & \xmark \\
		\texttt{06-Problems} & \cmark & \cmark & \xmark & \xmark \\
		\texttt{07-Products} & \cmark & \cmark & \cmark & \xmark \\
		\texttt{08-Widgets} & \cmark & \cmark & \cmark & \cmark \\
		\texttt{09-Safrs} & \cmark & \cmark & \cmark & \xmark \\
		\texttt{10-Realworld} & \cmark & \cmark & \cmark & \xmark \\
		\texttt{11-Crud} & \cmark & \cmark & \cmark & \xmark \\
		\texttt{12-Order} & \cmark & \cmark & \cmark & \xmark \\
		\texttt{13-Users} & \cmark & \cmark & \cmark & \xmark \\
		\texttt{14-Scheduler} & \xmark & \cmark & \xmark & \xmark \\
		\hline\hline
		\textbf{Total}: & \textbf{11} & \textbf{14} & \textbf{8} & \textbf{2}\\
		\hline
	\end{tabular}
	}
	\hspace*{15pt}
	\parbox[t]{.48\linewidth}{
	\caption{Coverage: number of ``wins'' for the tools on each metric.} 
	\label{tab:bestOf3}
	\begin{tabular}{l|ccc|c}
		\hline
		\textbf{Coverage metric} & \textbf{\rtg} & \textbf{\rler} & \textbf{\bb} & \textbf{Draw} \\ 
		\hline\hline
		Path & \tabhighlight{1} & 0 & 0 & 7 \\ 
		Operation & 1 & \tabhighlight{3} & 0 & 4 \\ 
		Parameter & \tabhighlight{1} & 0 & 0 & 4 \\ 
		Parameter value & 1 & 0 & \tabhighlight{2} & 0 \\ 
		Req. content-type & \tabhighlight{2} & 1 & 0 & 4 \\ 
		\hline
		Status code class & 4 & 4 & \tabhighlight[white]{0} & 0 \\ 
		Status code & \tabhighlight{5} & 3 & 0 & 0 \\ 
		Resp. content-type & \tabhighlight{3} & 2 & 0 & 2 \\ 
		\hline
	\end{tabular}
	}
\end{table}

\rler resulted the most robust automated testing tool because it is the only one able to manage all case studies. Indeed, all other tools fail (or crash) while testing some~services.

\rtg was the second most robust tool, as it could run on 11 case studies out of 14. During the testing of case study \texttt{02}, the tool got stuck in an endless loop while parsing a date. In case studies \texttt{03} and \texttt{14}, instead, the tool crashed in its initialization phase. \rtg makes use of the official \emph{Swagger Codegen}\cite{Codegen} to build HTTP client classes starting from the \openapi specification, and the failure is due to this module which is executed at the very beginning.

\bb could run on approximately half of the APIs (8 out of 14). In case study \texttt{01}, the tool crash is caused by an unhandled Java Null Pointer Exception of the component responsible to write the Excel output file. In the other 5 non-working cases, \bb crashes while parsing the \openapi specification, especially when resolving the defined schemas.

\rets seems to be the least robust testing tool because it failed on most of the case studies. Indeed it could only test two of them: \texttt{01} and \texttt{08}. The main limitation of the tool is its inability to test REST APIs that use body parameters (\eg a JSON data structure in the body) when no body parameter examples are provided within the specification. Nevertheless, this is a quite common scenario in practice: in fact, this happens for 11 out of 14 case studies. In one other case (case study \texttt{09}), the failure is due to malformed requests containing two \emph{content-type} fields in the header. When building requests, \rets sets the default content-type field in the header. A second content-type field is then appended, in case the specification explicitly documents it. However, the component in \rets responsible for checking the request correctness detects the duplicate content-type field in the header and stops the program execution with an error message. Basically, the tool rejects the request generated by the tool itself. 



Based on these results, we can answer RQ1 as follows:\\[-5mm]

\begin{center}
\fbox{
\begin{minipage}{.975\linewidth}
{\em
\rler is the most robust automated testing tool for REST APIs, because it could test all the 14 case studies. \rtg is the second most robust tool (11/14), followed by \bb (8/14). \rets is the least robust automated testing tool (2/14).
}
\end{minipage}
}
\end{center}

\vspace*{3pt}

\subsection{RQ2: Analysis of Coverage}

We now compare the automated testing tools with respect to the coverage metrics. We have excluded \rets from the comparison due to its very low robustness. Indeed, restricting the comparison on only the two case studies testable by \rets would have made the coverage analysis meaningless. Therefore, in order for the comparison to be as fair as possible, we considered only the case studies that all the remaining tools (\rtg, \rler and \bb) could test successfully. Thus, the coverage comparison focuses on 8 case studies, namely \texttt{05} and from \texttt{07} to \texttt{13}. 

\begin{figure*}[t]
	\centering
	\includegraphics[scale=0.65]{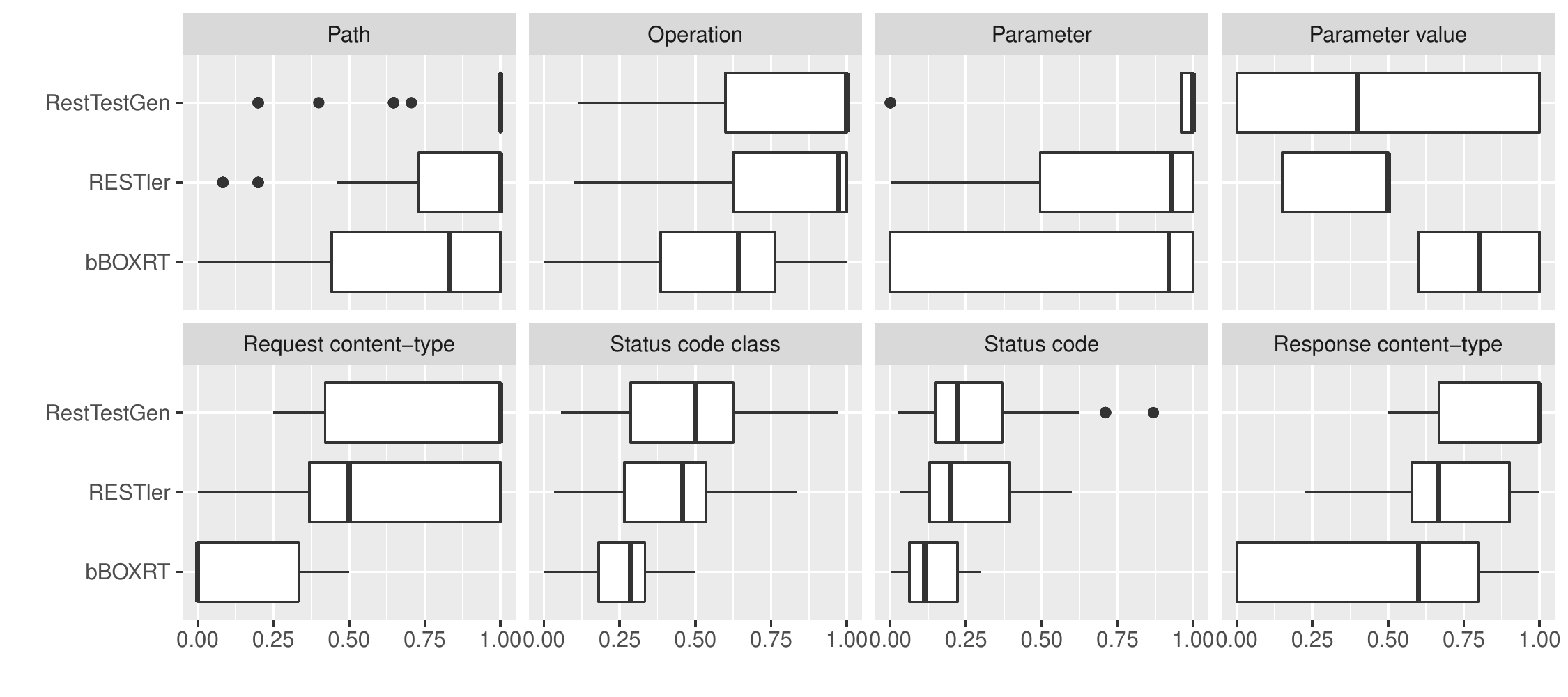}
	\vspace*{-15pt}
	\caption{Box-plots of coverage metrics on the 8 selected case studies.}
	\label{fig:metrics}
	
\end{figure*}

Figure~\ref{fig:metrics} shows the experimental data distributed over the 8 selected case studies, with a different box-plot for each metric. For instance, in the second box-plot of the first row we can compare values of the \emph{Operation} coverage metric. While \rtg and \rler score very high and similar values of Operation coverage, \bb records lower values.

Overall, \rtg seems superior to the other tools with respect to \emph{Parameter}, \emph{Request content-type}, \emph{Response content-type}, \emph{Status code class} and \emph{Status code} coverage metrics. \rtg and \rler achieve similar values of \emph{Path} coverage (they have the same median). Eventually, \bb overcomes the other testing tools on \emph{Parameter value} coverage. \rler does not perform better than the other tools with respect to any coverage metric. 


After these qualitative considerations on graph trends, we mean to present more quantitative comparison results. We recorded all the coverage metrics for all testing tools on all case studies, and we take the average value on 10 runs in order to avoid bias due to the nondeterministic components of the tools. We say that a testing tool hits a ``win'' for a coverage metric on a case study if the testing tool scores the highest value for that metric when testing that case study. 


Table~\ref{tab:bestOf3} reports the number of ``wins'' for each testing tool on each coverage metric.
For instance, the second line shows the results for the \emph{Operation} coverage metric. We can observe that \rtg reported higher Operation coverage than \rler and \bb on 1 case study. Instead, \rler reported the highest value of Operation coverage on 3 case studies. \bb never reported a higher value for this metric. For the remaining 4 case studies no testing tool is a clear winner, because at least two other tools reported an equally high value (\emph{Draw} column). Thus, we can claim that \rler is preferable when considering Operation coverage (the corresponding number of ``win'' is highlighted in the~table). 

Not all rows of Table~\ref{tab:bestOf3} add up to 8, since a metric may not be computable for a specific case study. For instance, \emph{Request content-type} coverage can be computed only when operations content-types have no wildcard (see Section~\ref{sec:metrics} for details).

According to these data, \rtg is preferable for \emph{Path}, \emph{Parameter}, \emph{Request content-type}
\emph{Status code} and \emph{Response content-type} coverage. 
\rler recorded the best results for \emph{Operation} coverage, while \bb for \emph{Parameter value} coverage. 
\rtg and \rler achieved equally high results (4 case studies) for \emph{Status code class} coverage.


With these results, we can answer RQ2 as follows:\\[-5mm]
\begin{center}
\fbox{
\begin{minipage}{.975\linewidth}
{\em
\rtg is the automated testing tool producing test suites with higher coverage, because when generating test cases for 8 case studies it overtakes the other testing tools on 5 coverage metrics, while \rler and \bb have been superior to the other tools according to 1 coverage metric each. 
}
\end{minipage}
}
\end{center}

\vspace*{4pt}

\subsection{Considerations}

Based on the experimental results, we could formulate the following subjective considerations. 

\fakeparagraph{Research prototypes robustness} 
The four tools under analysis are all research prototypes, so, it is not surprising that they may fail in testing some real-world case studies. Nevertheless, \rler is the most mature tool, able to test without errors all the considered APIs. The different robustness degrees among the tools poses a first obstacle to the comparison of coverages. Indeed, while we were able to compare \rler, \rtg and \bb on a sufficiently large set of case studies, a comparison between all four tools would have led to only 2 common case studies, resulting in a not effective comparison. 

\fakeparagraph{Enumeration affects testing budget} Each tool that we considered in our empirical comparison adopts a distinct approach for assembling sequences of operations into test scenarios. 
An exhaustive enumeration of sequences (\rler) seems to be very time-consuming and less time-effective than a well-thought-out single sequence (\rtg). This highlights that when a large amount of requests (and consequently a large amount of time) cannot be spent in testing, focused approaches (such as the one adopted by \rtg) are preferable. Conversely, when a lot of resources can be allocated to testing, possibly in the cloud, testing an exhaustive enumeration of interaction sequences is probably acceptable.


\fakeparagraph{Input generation Vs sequence enumeration}
When generating test cases, search budget might be optimized across different dimensions, either to test with many different interaction sequences, or focusing on few sequences but with many different input data. According to our empirical observations, both of these dimensions are important, but, especially when testing time is limited, focusing on exploring new input data (\rtg) seems to achieve higher generic coverage than exploring new interaction sequences (\rler). 

\fakeparagraph{Multiple input generation algorithms}
When generating test input data, some approaches focus on single strategies (such as using random values, dictionaries, mutations, etc.) or a combination of these. However, rather than just adopting a few data generation strategies, we observed that higher generic test coverage is achieved when more of those strategies are integrated and combined, including also data that have been observed as output of previously executed tests. In fact, test outputs might represent valid actual data from the database of the REST API under test. Investigating novel input data generation strategies is a promising research direction to deliver more effective testing approaches.



\fakeparagraph{Coverage according to input Vs output metrics} 
High coverage with respect to \emph{input} metrics seems to be easier to obtain, rather than high coverage with respect to \emph{output} metrics. In fact, each tool scored very high values of {\em Operation}, {\em Path} and {\em Parameter} coverage, which are all input metrics. Indeed, obtaining 100\% coverage for these input metrics is relatively easy: it just requires a tool to exercise once those operations, paths and input parameters that are documented in the \openapi specification. 


Conversely, obtaining 100\% coverage for output metrics is more challenging. While requests and input data are selected by testing tools, responses and output data are not directly under control of testing tools. For instance, a tool can not simply decide to cause a response with either a {\em correct} status code (\eg \twoxx) or an {\em error} status code (\eg \fivexx). To obtain distinct status codes, a testing tool has to guess both valid and invalid input data, which is challenging, and it might require several attempts. Hence, output coverage metrics can be considered harder to satisfy, because they require out-of-specification knowledge about the service under test, and they can be considered a more reliable indicator of deeper testing than input coverage metrics.

\section{Related Work}
\label{sec:related}


Existing commercial test authoring tools, like~\cite{Postman,Soap,vREST,httpMaster,APIFortress,RESTAssured}, help developers to \emph{manually} write tests that can be then automatically run by the tool. These approaches are not fully automatic, as the tools we have considered in the present work. 

Concerning automatic tests generation for REST APIs, the research community proposes some interesting solutions, following mainly two different lines of work. One consists in \emph{white-box} approaches, that rely on the availability of APIs source code to perform static analysis, or to instrument it to collect execution traces and metric values. In this context, Arcuri~\cite{Arcuri2019EvoMaster} proposes a fully automated solution to generate test cases with evolutionary algorithms, that requires the \openapi specification and the access to the Java bytecode of the REST API to test. This approach has been implemented as a tool prototype called EvoMaster, extended with the introduction of a series of novel testability transformations aimed at providing guidance in the context of commonly used API calls~\cite{Arcuri2020TestTransformations}.

On a complementary direction, \emph{black-box} approaches do not require any source code, which is often the case when using closed-source components and libraries. \emph{Fuzzers}~\cite{APIFuzzer,FuzzLightyear,FuzzySwagger,SwaggerFuzzer,TnTFuzzer} are black-box testing tools that generate new tests starting from previously recorded API traffic: they fuzz and replay new traffic in order to find bugs. Some of these also exploit the \openapi specification of the service under test~\cite{FuzzLightyear,FuzzySwagger,SwaggerFuzzer}. Godefroid et al.~\cite{Godefroid2020Fuzzing} propose a methodology to fuzz body payloads intelligently using JSON body schemas and advanced fuzzing rules 
(as done in RESTler~\cite{Atlidakis2019RESTler}). Although they are automatic black-box tools, their goal is to generate input values to tests, so they cannot be used as standalone testing tools (except for the approach of Godefroid et al.~\cite{Godefroid2020Fuzzing} that has been implemented in \rler).
Ed-douibi et al.~\cite{EdDouibi2018AutomaticTestsGen} propose a model-based approach for black-box automatic test case generation of REST APIs. A model is extracted from the \openapi specification of a REST API, to generate both nominal test cases (with input values that match the model) and faulty test cases (with input values that violate the model). However, we did not manage to install their proof-of-concept implementation (due to some errors in the source code), hence we had to exclude the work from our comparison.
Karlsson et al.~\cite{Karlsson2020QuickREST} propose QuickREST, a tool for property-based testing of RESTful APIs. Starting from the \openapi specification, they generate test cases with the aim of verifying whether the API under test complies with some properties (\ie definitions) documented in the specification (\eg status codes or schemas). Unfortunately, we had to exclude QuickREST from our work because the version we found online was incompatible with the case studies we randomly selected for our comparison (it did not manage to test any of our case studies). 
Segura et al.~\cite{Segura2018MetamorphicRelations} propose another black-box approach, where the oracle is based on metamorphic relations among requests and responses. For instance, they send two queries to the same REST API, where the second query has stricter conditions than the first one (\eg by adding a constraint). The result of the second query should be a proper subset of entries in the result of the first query. When the result is not a sub-set, the oracle reveals a defect. However, this approach only works for search-oriented APIs. Moreover, this technique is only partially automatic, because the user is supposed to manually identify the metamorphic relation to exploit and  what input parameters to test.

To the best of our knowledge, the only black-box testing approaches for REST APIs which provide an implementation, \ie a usable testing tool, are the one we have taken into account in our comparison (\rtg~\cite{Viglianisi2020RESTTESTGEN}, \rler~\cite{Atlidakis2019RESTler}, \bb~\cite{Laranjeiro2021BBOXRT} and \rets~\cite{MartinLopez2020RESTest}, presented in Section~\ref{sec:objTools}). Regarding test coverage, the only work proposing a systematic approach to assess the coverage of REST APIs testing tools is the framework of Martin-Lopez et al.~\cite{MartinLopez2019TestCoverage}, that we have taken as basis for our comparison.

\section{Conclusion}
\label{sec:conclusion}

Despite several approaches and automated tools are available to test cases generation for REST APIs, the literature is still missing an explicit comparison of them. In this paper, we defined an experimental framework that includes a benchmark of REST APIs case studies and a coverage measurement infrastructure. 
We adopted this framework to carry out a comparison of four state-of-the-art automated black-box REST APIs testing tools (\rtg, \rler, \bb, and \rets) in terms of robustness and~coverage.

\rler appears to be the most robust tool, being able to test all the case studies without incurring in crashes or failures. Instead, the strategy of \rtg, based on data dependencies among operations, appears to be the most effective, as it overtakes the other approaches in several coverage metrics.

Based on our experimental results, we formulated some considerations that might guide developers in making an informed decision on which tool to adopt.

As a future work, we plan to evolve the comparison with new testing tools, along with the updated versions of the already considered tools. Another interesting aspect we plan to investigate is how the specification-based coverage we adopted correlates with code coverage. Indeed, even if code coverage is inevitably more accurate, sometimes a ``black-box coverage'' is the only viable option. With a study correlating the two approaches we can asses in which situations, or under which assumptions, the specification-based coverage is  still an acceptable solution. Finally, we intend to extend our experimental framework, that currently just focuses on test coverage, to also measure defect detection capabilities.

\section*{Acknowledgment}
This paper has been partially supported by project MIUR 2018-2022 ``Dipartimenti di Eccellenza''.

\bibliographystyle{IEEEtran}
\bibliography{bib}

\end{document}